\documentclass[sigplan,screen]{acmart}    

\usepackage{newunicodechar}

\AtBeginDocument{%
  }

\setcopyright{acmlicensed}
\copyrightyear{2026}
\acmYear{2026}
\acmDOI{XXXXXXX.XXXXXXX}
\acmConference[SLE '26]{ACM SIGPLAN International Conference on Software Language Engineering (SLE)}{July 02--03,
  2026}{Rennes, France}

\usepackage{proof}
\usepackage[ruled]{algorithm2e}
\usepackage{algpseudocode}
\usepackage{color}
\usepackage{amsmath}
\usepackage{acronym}
\usepackage{listings}
\usepackage{url}

\newcommand{\term}[1]	{\emph{#1}}

\newcommand{\code}[1]{\texttt{#1}}

\newcommand{\idx}[2][]	{\ifthenelse{\equal{#1}{}}{\index{#2}#2}{\index{#1}#2}}

\newcommand{\eaci}[1]{\edef\tmp{\noexpand\index{\ac{#1}}} \emph{\tmp}}

\newcommand{\defvoc}[2]{\expandafter\def\csname #1\endcsname{#2} {\newacro{#1}{#2}}}

\newcommand{\vf}[1]{\def\tmp{\index{\csname #1\endcsname@{\acf{#1}}}}\tmp\emph{\acf{#1}}}
\newcommand{\vs}[1]{\def\tmp{\index{\csname #1\endcsname@{\acf{#1}}}}\tmp\acs{#1}}
\newcommand{\vl}[1]{\def\tmp{\index{\csname #1\endcsname@{\acl{#1}}}}\tmp\acl{#1}}

\definecolor{keywordcolor}{rgb}{0.7, 0.1, 0.1}   
\definecolor{tacticcolor}{rgb}{0.0, 0.1, 0.6}    
\definecolor{commentcolor}{rgb}{0.4, 0.4, 0.4}   
\definecolor{symbolcolor}{rgb}{0.0, 0.1, 0.6}    
\definecolor{sortcolor}{rgb}{0.1, 0.5, 0.1}      
\definecolor{attributecolor}{rgb}{0.7, 0.1, 0.1} 

\newunicodechar{→}{\ensuremath{\to}}
\newunicodechar{ℝ}{\ensuremath{\mathbb{R}}}
\newunicodechar{𝕜}{\ensuremath{\mathbb{k}}}
\newunicodechar{𝓝}{\ensuremath{\mathcal{N}}}
\newunicodechar{∀}{\ensuremath{\forall}}
\newunicodechar{∃}{\ensuremath{\exists}}
\newunicodechar{ℤ}{\ensuremath{\mathbb{Z}}}
\newunicodechar{λ}{\ensuremath{\lambda}}
\newunicodechar{α}{\ensuremath{\alpha}}
\newunicodechar{β}{\ensuremath{\beta}}
\newunicodechar{∘}{\ensuremath{\circ}}
\newunicodechar{≠}{\ensuremath{\neq}}
\newunicodechar{↦}{\ensuremath{\mapsto}}
\newunicodechar{∧}{\ensuremath{\wedge}}
\newunicodechar{₁}{\ensuremath{_1}}
\newunicodechar{₂}{\ensuremath{_2}}
\newunicodechar{₁}{\ensuremath{_1}}
\newunicodechar{₃}{\ensuremath{_3}}
\newunicodechar{¹}{\ensuremath{^1}}
\newunicodechar{⁻}{\ensuremath{^-}}
\newunicodechar{≤}{\ensuremath{\le}}
\newunicodechar{≥}{\ensuremath{\ge}}

\begin{document}
\def\lstlanguagefiles{lstlean.tex, lstsouffle.tex}
\lstset{language=lean}

\title{A Shallow Embedding of Datalog in Lean}
\author{Ramy Shahin}
\orcid{0000-0001-8724-3934}{}
\affiliation{
  \institution{Qualgebra}
  \country{Canada}
}
\email{ramy@qualgebra.com}

\begin{abstract}
Datalog is a lightweight logic programming language, based on the logic of Horn clauses. 
Lean, on the other hand, is a proof assistant system and language based on the Calculus of Inductive Constructions (CIC). 
Datalog is more constrained and less expressive than Lean but has a long history of established deduction algorithms. 
Writing definitions and queries in the Datalog fragment of Lean would be more succinct and understandable than writing them in Lean itself.

This paper outlines the design and implementation of a shallow embedding of Datalog as a Domain Specific Language (DSL) on top of Lean. 
Bidirectional interoperability between the Datalog DSL and Lean is a primary goal of this design. 
In addition to rules and facts, backward chaining queries are automatically translated into theorems with tactic-based proofs. 
The paper also includes three simple examples of how the DSL can be used.
\end{abstract}



\maketitle

\section{Introduction}
\label{sec:intro}

Logical inference is the process of deducing new facts from known ones.
Given the pair of facts (1) \term{Socrates is a man}, and (2) \term{all men are mortal}, we can infer that (3) \term{Socrates is mortal}.
To reach this conclusion, we are implicitly using the classical logic principle of \term{modus ponens}.
Many modern computer applications employ this kind of inference to help analyze extensive bodies of facts and to help in decision-making.

Datalog~\cite{Ceri:1989,Ceri:1990} is a logic programming language allowing users to encode known facts and inference rule.
\term{Forward-chaining} is a \term{bottom-up reasoning} algorithm, recursively generating new facts inferred from known facts and inference rules.
If all data domains are finite (e.g., bounded integers, finite sets of strings), this algorithm is guaranteed to terminate in polynomial time with respect to the number of input facts~\cite{Ceri:1989}. 

On the other hand, in contexts where inferrable facts are known to be infinite, or if only a limited number of facts is of interest, users can specify which facts are to be provably inferred, and the Datalog engine tries to construct a \term{proof tree} for each of those facts. 
The proof tree is constructed from input facts and rules.
If the engine fails to construct a proof tree for a fact, then either this fact is not a logical consequence of the inputs, or the engine ran out of computational resources trying to construct the tree.
This \term{top-down} approach is referred to as \term{backward-chaining} because it starts with the goal at the root of the tree, and tries to build the rest of it all the way down to an input fact in each of the leaves.

Another class of logic-based systems and languages is Interactive Theorem Provers (ITPs).
An ITP allows users to write definitions, theorems, and proofs of those theorems.
The ITP automatically checks the logical \term{validity} of those proofs, and does not admit a theorem without a valid proof.
Examples of ITPs include Coq~\cite{Bertot:2010}, Isabelle/HOL~\cite{Nipkow:2002}, and Lean~\cite{Moura:2021}.

Datalog and ITPs clearly have different use cases.
Inference is automatic in Datalog, while in ITPs the user has to provide a valid proof for any new fact.
On the other hand ITPs are much more expressive and flexible, and the syntax of their languages and their type systems allow for much richer definitions and theorems not expressible in Datalog. 

This paper presents the design and implementation of a Datalog Domain Specific Language (DSL), embedded into the Lean language.
Embedded DSLs are language extensions attached to host languages.
This allows users to write programs otherwise not easily written in the host language by itself.
The host language in this case is Lean, and the DSL includes syntactic constructs for Datalog facts, rules, and some constructs facilitating the interoperability between the DSL and Lean.
Lean includes a set of metaprogramming features that make the development of this DSL less complicated, compared to other host languages that do not support metaprogramming or syntax extensions.

Because Datalog constructs in the DSL are translated into native Lean constructs, this is considered a \term{shallow embedding} of Datalog into Lean.
In particular, Datalog facts and rules are translated into Lean axioms, Datalog queries are translated into Lean theorems, and proof trees are encoded into Lean proof terms.
In addition, the DSL does not define new types, so it would not violate any of the type-theoretic restrictions of Lean (e.g., positivity checking of recursive types~\cite{Abbott:2005}).

An underlying assumption of this work is that using Datalog together with Lean makes sense mostly in the context of automatically constructing certified proofs of specific goals, as opposed to enumerating all inferrable facts.
Given this assumption, the DSL design covers only backward-chaining inference.
This and other design decisions aim at fulfilling a particular set of goals:
\begin{enumerate}
    \item To allow Datalog programmers not fully familiar with Lean to use Datalog fact and rule syntax in Lean code.
    \item To implement Datalog queries as Lean theorems, and to automatically construct Lean-checked proof terms of them using backward-chaining.
    \item To ensure bidirectional interoperability with Lean definitions, theorems, and proofs. Particularly, to allow the use of existing Lean theorems as Datalog rules, to use the results of Datalog queries as Lean theorems, and to allow the use of arbitrary Lean value types in Datalog atoms. 
\end{enumerate}

This paper makes the following contributions:
\begin{enumerate}
    \item Outlining the design of a Datalog DSL embedded in Lean (Sec.~\ref{sec:design}).
    \item Discussing the open-source implementation\footnote{\url{https://github.com/qualgebra/LDatalog/tree/SLE2026}} 
    of this DSL using Lean metaprogramming.
    \item Demonstrating the use of the DSL by three simple examples covering both inference and calculation of values satisfying Datalog queries (Sec.~\ref{sec:examples}). These examples also illustrate the interoperability of Datalog and Lean.
\end{enumerate}
\section{Background}
\label{sec:background}

\newcommand{\souffle}{Souffl\'{e}}

This section summarizes aspects of Datalog, Lean, and Lean metaprogramming on which the rest of the paper depends.

\subsection{Datalog}

\begin{figure}[t]
\begin{lstlisting}[language=souffle,xleftmargin=10pt,numbers=left]
.decl edge(from:symbol, to:symbol)
.decl path(start:symbol, end:symbol)

edge("a", "b").
edge("b", "c").
edge("b", "d").

path(x, y) :- edge(x, y).
path(x, z) :- path(x, y), edge(y, z).

.output path
\end{lstlisting}
\caption{Graph reachability example in the \souffle~Datalog syntax.}
\label{fig:datalogExample}
\end{figure}

Datalog is a declarative logic programming language combining the data definition features of SQL and some logical inference features of Prolog~\cite{Ceri:1989}. 
A Datalog program is composed of relation definitions, a set of logical facts, and a set of inference rules.
For example, the Datalog program in Fig.~\ref{fig:datalogExample} (in the syntax of the \souffle~Datalog engine~\cite{Jordan:2016}) encodes a directed graph in terms of its node labels as strings, and edges between nodes.

The example starts with the declaration of two relations: \code{edge} and \code{path} (lines 1-2).
The \code{edge} relation has two fields: \code{from}, the source of a directed edge in the graph, and \code{to}, the target node of the edge.
In this example a node is identified by a string label (the \code{symbol} type in \souffle).
Similarly, the \code{path} relation has two fields: \code{start}, the starting point of the path, and \code{end}, its ending point.

Three facts are then added to the database (lines 4-6).
Each of those facts add a record to the \code{edge} relation.
Lines 8-9 implicitly define the records of the \code{path} relation using a recursive pair of inference rules.
The base case (line 8) states that whenever there is an \code{edge} record between nodes \code{x} and \code{y}, there is a \code{path} between those two nodes.
The left-hand side of the rule is referred to as the rule \term{head}. 
The right-hand side, which is a comma-separated sequence of logical conjuncts, is referred to as the rule \term{body}.

The second rule in line 9 is the recursive case.
It states that whenever there is a \code{path} from \code{x} to \code{y}, and an \code{edge} from \code{y} to \code{z}, then there is a \code{path} from \code{x} to \code{z}.
The body in this rule consists of two \term{atoms}, and the rule is syntactically equivalent to an implication from the conjunction of the atoms in the rule body, to the rule head.

Line 11 instructs the \souffle~Datalog engine to write the contents of the \code{path} relation to a Comma-Separated Values (CSV) file. The contents of that file for this example are:
\begin{lstlisting}[language=souffle,xleftmargin=20pt]
a	b
a	c
a	d
b	c
b	d
\end{lstlisting}

\souffle~populates relations using the \term{forward-chaining algorithm}.
The algorithm loops over the rules of a Datalog program, and whenever the body of a rule can be satisfied by existing records, the head is added as a new fact to the database if it doesn't already exist. 
This is repeated until a fixed point is reached, i.e. no new facts are inferred.
Because the set of field values ("a", "b", "c", and "d" in this example) stays the same, the number of possible records for each relation is always finite.
As a result, the forward-chaining algorithm is guaranteed to terminate in polynomial time of the number of input facts~\cite{Ceri:1989}. 

Another approach to evaluating Datalog programs starts from a \term{goal} atom, and tries to build a \term{proof tree} of that goal using the facts and rules of the Datalog program~\cite{Ceri:1990}.
The goal is written as a Datalog atom ending with a question mark.
This is typically called a Datalog \term{query}.
For example, a query trying to prove the existence of a path from node \code{"a"} to node \code{"d"} is written as \code{path("a", "d")?}.

This top-down approach to evaluating Datalog queries is called backward-chaining, because it builds a proof backwards from the goal.
Backward-chaining is used in some Datalog engines, as well as in Prolog inference.
It is better suited in settings where the number of field values is not necessarily finite.
For example, if the language supports mathematical operations on unbounded natural numbers, a base fact and a rule such as
\begin{lstlisting}
    R(0).
    R(x + 1) :- R(x).
\end{lstlisting}
would generate an infinite inductive unary relation \code{R}, with the base case \code{R(0)}, and an inductive rule \code{R(x) $\implies$ R(x+1)}.
This inductive definition includes every natural number in relation \code{R}.
As a result forward-chaining (enumerating all members of \code{R}) would not terminate.
Backward-chaining on the other hand can still construct a finite proof tree for a specific goal, such as \code{R(77)?}.

This paper only covers backward-chaining because Lean supports infinite types, and backward-chaining can be used as a proof search technique together with other Lean tactics.

Backward-chaining is a backtracking algorithm.
It starts with the query goal, and checks whether it unifies with the head of any of the program rules.
If it does, the atoms of the body of that rule are added as new proof goals.
The process is repeated for any remaining proof goals, until they are all discharged.
If the algorithm is stuck (cannot find a rule whose head unifies with the current goal), it backtracks, and tries a different rule in the previous step.
The algorithm is typically parametrized by a maximum backtracking depth to make sure it doesn't recurse forever.

\subsection{Lean}

\begin{figure}[t]
\begin{lstlisting}[xleftmargin=10pt,numbers=left]
inductive edge: String → String → Prop
| f₁: edge "a" "b"
| f₂: edge "b" "c"
| f₃: edge "b" "d"

inductive path: String → String → Prop
| r₁: edge x y → path x y
| r₂: path x y → edge y z → path x z

theorem th₁: path "a" "d" := 
   path.r₂ (path.r₁ edge.f₁) edge.f₃
\end{lstlisting}
\caption{Graph reachability example in Lean.}
\label{fig:leanExample}
\end{figure}

Lean is both an Interactive Theorem Prover (ITP) and a programming language~\cite{Moura:2021}.
Users can write definitions, theorems, proofs, functional programs, and metaprograms (programs manipulating other Lean programs) all in the same language.
The underlying core of Lean is the Calculus of Inductive Constructions (CIC)~\cite{Bertot:2010,Pfenning:1990}.
According to the Curry-Howard correspondence, theorems in Lean are types, and their proofs are terms of those types.

Fig.~\ref{fig:leanExample} is a Lean encoding of the directed graph example in Fig.~\ref{fig:datalogExample}.
The \code{edge} relation (together with its records) is encoded as an inductive binary predicate (lines 1-4). The predicate has two parameters of type \code{String} (the two end-points of the directed edge), and its return type is \code{Prop} (for propositional), which indicates that this is a logical definition, not a computable function.

The three edge facts in the Datalog example are implemented as the three constructors \code{f}$_1$, \code{f}$_2$, and \code{f}$_3$ of the inductive predicate \code{edge}.
The \code{path} inductive predicate type (lines 6-8) has two constructors: \code{r}$_1$ corresponds to the base case rule of the Datalog \code{path} relation, while \code{r}$_2$ is the recursive case.

To assert any property on this directed graph, the property is formulated as a theorem, and a proof has to be provided for this theorem.
For example, to prove that there is a path from node \code{"a"} to node \code{"d"}, theorem \code{th$_1$} of type \code{path "a" "d"} is defined (line 10).
Lean only admits a theorem only if its proof is provided.
The proof of \code{th$_1$} is the term \code{path.r$_2$ (path.r$_1$ edge.f$_1$) edge.f$_3$} (line 11).

Mathlib~\cite{Mathlib:2020} is an extensive library of formalized mathematics written in Lean.
It includes definitions and theorems in various fields of mathematics.
The example in Sec.~\ref{sec:derivatives} utilizes some Mathlib calculus theorems to demonstrate how the Lean Datalog DSL can be used to compute function derivatives, together with a correctness proof.

\subsection{Lean Metaprogramming}

Starting with Lean 4~\cite{Moura:2021}, a highly expressive metaprogramming framework written in Lean itself is a part of the language.
This framework allows the development of proof tactics (proof automation programs), syntax extensions, environment extensions, and Domain Specific Languages (DSLs) in Lean.

Lean tactics are metaprograms manipulating the internal state of the current proof, trying to synthesize Lean terms that would discharge pending proof obligations.
Tactics typically capture reusable and commonly-used proof patterns, significantly simplifying proof writing in many cases.
Lean comes with a set of standard tactics, and Mathlib includes an even wider set. In addition, end users can always write their own.

Syntax extensions come in the form of user-defined syntactic categories with grammar rules.
Lean generates parsers based on the grammar rules, registers those parsers, and automatically invokes them when the syntax extension is used in a Lean program.
Parsing turns sequences of tokens into \term{terms}.
When developing a syntax extension, a user-defined \term{elaborator} translates terms into Lean expressions.
Those expressions are then compiled by the Lean compiler into the minimalistic kernel language, which is mostly low-level CIC constructs.
During compilation, the Lean type checker validates the logical soundness of proofs written in the source language.

Embedding a DSL in Lean is effectively the process of designing a Lean syntax extension, together with its set of elaborators.
For example, embedding the syntax of Datalog facts and rules in Lean would involve two new syntactic categories, one for facts and the other for rules.
The elaborators of those new syntactic categories would generate plain Lean definitions that correspond to the input facts and rules. 

\section{DSL Design}
\label{sec:design}

\newcommand{\lterm}{\text{<term>}}
\newcommand{\lident}{\text{<ident>}}
\newcommand{\args}{\text{<Args>}}
\newcommand{\arglist}{\text{<ArgList>}}
\newcommand{\atom}{\text{<Atom>}}
\newcommand{\atomList}{\text{<AtomList>}}
\newcommand{\fact}{\text{<Fact>}}
\newcommand{\query}{\text{<Query>}}
\newcommand{\drule}{\text{<Rule>}}
\newcommand{\statement}{\text{<Statement>}}
\newcommand{\identList}{\text{<identList>}}

This section outlines the design of the Lean Datalog DSL.
This includes the syntax of the DSL, the syntactic rewrite rules from DSL constructs to Lean constructs, and how Datalog queries are evaluated.

\subsection{Syntax}
\label{sec:syntax}

\begin{figure}[t]
\begin{center}
\[
\begin{array}{lcl}
      \args & ::= & \lterm \\
            & |   & \lterm~\code{,}~\args \\
      \arglist & ::= & \code{()} \\
               & |   & \code{(} \args \code{)} \\
      \atom & ::= & \lident \arglist \\
            & |   & \code{(}\lterm\code{)} \\
      \fact & ::= & \atom~\code{.} \\
      \query & ::= & \atom~\code{?} \\
      \atomList & ::= & \atom \\
                & |   & \atom~\code{,}~\atomList \\
      \drule & ::= & \atom~\code{:-}~\atomList~\code{.} \\
      \identList & ::= & \lident \\
                 & |   & \lident~\code{,}~\identList \\
      \statement & ::= & \fact \\
                 & |   & \query \\
                 & |   & \drule \\
                 & |   & \lident~\code{:}~\fact \\
                 & |   & \lident~\code{:}~\query \\
                 & |   & \lident~\code{:}~\drule \\
                 & |   & \code{use}~\identList~\code{.} \\
\end{array}
\]
\end{center}
\caption{Grammar of the Lean DSL.}
\label{fig:grammar}
\end{figure}

To facilitate the interoperability between the Datalog DSL and Lean, the syntactic category of Lean terms (\lterm) is a building block of the DSL grammar (Fig.~\ref{fig:grammar}).
A sequence of arguments to a Datalog rule or fact (\args) is a comma-separated sequence of Lean terms.

An argument list \arglist~encloses \args~(or an empty list) between parentheses.
A Datalog atom (\atom) is either a Lean identifier followed by an \arglist, or a Lean term (\lterm) enclosed within parentheses.
The former is standard Datalog syntax, while the latter is syntax added to the DSL in order to allow using arbitrary Lean terms as Datalog atoms. 
This is particularly useful when using built-in Lean syntax, such as infix operators (an example can be found in Sec.~\ref{sec:recOverlap}).

A Datalog fact (\fact) is syntactically an atom followed by a '.'.
Similarly, a Datalog query (\query) is an atom followed by a '?'.
A list of atoms (\atomList) is a comma-separated sequence of atoms.
A Datalog rule (\drule) is a single \atom~(head) and an \atomList~(body).
The head and the body are separated by ':-', and rule definition ends with '.'.

A list of identifiers (\identList) is a comma-separated sequence of Lean identifiers. 
A DSL statement (\statement) is a fact, query, rule, or a \code{use} statement. 
Facts, queries, and rules can also be prefixed with a label (a Lean identifier followed by ':'). 
A \code{use} statement specifies a list of Lean identifiers, each of which labeling a Lean theorem.
Those theorems will be used by the backward chaining algorithm when evaluating queries.

\subsection{Rewrite Rules}
\label{sec:rules}

\begin{table*}[t]
\centering

\begin{tabular}{|c|c|c|}
\hline
Syntactic Category & Expansion & Generated Lean Code \\
\hline
\hline
\atom & $(t)$ & $t$   \\
\cline{2-3}
      & $i~(a_1, ..., a_n)$ & $i~a_1~...~a_n$ \\
\hline
\fact & $a.$ & rewrite($a$) \\
\hline
\drule & $h$ :- $b_1,...,b_n$. & rewrite($b_1$) $\to$ ... $\to$ rewrite($b_n$) $\to$ rewrite($h$) \\
\hline
\query & $q$? & rewrite($q$) \\
       &      & (replacing placeholder variables with fresh Lean metavariables) \\
\hline
\statement & \code{use} $th_1, ..., th_n$. & \code{attribute [local datalog\_rule]} $th_1$ \\
           &                               & ... \\
           &                               & \code{attribute [local datalog\_rule] $th_n$}\\
\cline{2-3}
           & $f.$ & \code{@[datalog\_rule] axiom}: rewrite($f$) \\
\cline{2-3}
           & $r.$ & \code{@[datalog\_rule] axiom}: rewrite($r$) \\
\cline{2-3}
           & $q?$ & \code{@[datalog\_rule] theorem}: rewrite($q$) := \\
           &      & \code{by intros; solve\_by\_elim using datalog\_rule} \\
\cline{2-3}
           & $l: f.$ & \code{@[datalog\_rule] axiom} $l$: rewrite($f$) \\
\cline{2-3}
           & $l: r.$ & \code{@[datalog\_rule] axiom} $l$: rewrite($r$) \\
\cline{2-3}
           & $l: q?$ & \code{@[datalog\_rule] theorem} $l$: rewrite($q$) := \\
           &         & \code{by intros; solve\_by\_elim using datalog\_rule} \\
\hline
\end{tabular}

\caption{Rewrite rules from the Datalog DSL to plain Lean.}
\label{fig:rules}
\end{table*}

Table~\ref{fig:rules} outlines the rewrite rules from the Datalog syntactic constructs (Sec.~\ref{sec:syntax}) to plain Lean code. 
This is how the semantics of the DSL are embedded in Lean, and how Datalog sections can interoperate with Lean code.
These rules are implemented by the \code{rewrite} function, which is called recursively on the syntactic structure of a DSL fragment.

The \atom~syntactic category has two forms: a Lean term $t$ enclosed in parentheses, which is rewritten simply into $t$, or an identifier $i$ followed by a comma-separated argument list $(a_1,...,a_n)$, which is rewritten into the function application syntax of Lean $i~a_1~...~a_n$.

A fact is syntactically just a single atom, so rewriting a fact composed of atom $a$ is simply applying previous rewrite rules to $a$.
A rule composed of a head atom $h$ and a list of body atoms $b_1$,...,$b_n$ is rewritten into a function type with rewrite($b_1$), ..., rewrite($b_n$) as the types of the function parameters, and rewrite($h$) as the return type.

Similar to facts, a query $q$ is a single atom, so it is rewritten into rewrite($q$).
However, some of the query parameters might be placeholder variables (ending with '?')\footnote{An identifier ending with '?' is a valid Lean identifier.}.
When the query atom is rewritten, those placeholders are replaced with a fresh Lean metavariable each.
When proving the theorem corresponding to the query, the unification algorithm (part of the Lean type system) instantiates those metavariables.

For example, applied to the graph encoded in Fig.~\ref{fig:leanExample}, a query attempting to find a node reachable from \code{"b"} is written in the DSL as follows:
\begin{lstlisting}
      q: path("b", m?)?
\end{lstlisting}
The placeholder \code{m?} (representing the end point of a path starting at \code"b") should be instantiated as a part of evaluating this query.
This DSL query is translated into this Lean theorem definition:
\begin{lstlisting}
      theorem q: path "b" ?m
\end{lstlisting}
The Lean metavariable \code{?m} is different from the DSL placeholder \code{m?}, but they are in 1-1 correspondence to make sure that multiple occurrences of the same placeholder in a DSL query map to the same Lean metavariable in a theorem definition.
Proving this theorem in Lean involves instantiating \code{?m} with a concrete satisfiable value.
There are two facts in this example (together with rule \code{path.r$_1$}) that would prove this theorem: \code{edge.f$_2$} would instantiate \code{?m} with the value \code{"c"}, while \code{edge.f$_3$} would instantiate it with \code{"d"}.

A \code{use} statement takes a list of Lean theorem names. 
Those theorems will be used by the backward chaining algorithm when evaluating Datalog queries.
To distinguish those theorems from all the other Lean theorems already in the current context, the \code{use} $th_1, ..., th_n.$ statement generates a sequence of Lean attribute declarations, attaching the \code{datalog\_rule} attribute to each of those theorems. 
Syntactic attributes in Lean annotate definitions with attribute tags, and Lean metaprograms can distinguish definitions with those annotations and provide special processing logic for them.
This way the backward chaining algorithm (Sec.~\ref{sec:backChain}) specifically uses those theorems, together with the Datalog facts and rules.

A fact statement $f$ is rewritten into a Lean axiom of type rewrite($f$).
Similarly, a rule statement $r$ is rewritten into an axiom of type rewrite($r$).
A query $q$ is rewritten into a theorem of type rewrite($q$).
This theorem type includes a Lean metavariable for each placeholder variable in $q$.
The value assigned to this theorem (its proof) is calculated using the backward chaining algorithm discussed in Sec.~\ref{sec:backChain}. 
If a fact, rule, or query is prefixed by a label $l$, $l$ becomes the name of the axiom (theorem in the case of queries), otherwise it becomes anonymous (Lean assigns it a system-defined name).
 
Definitions of axioms for facts and rules, and theorems for queries, are all prefixed by the \code{datalog\_rule} attribute. 
This marks those axioms and theorems as definitions of interest to the backward chaining algorithm when evaluating queries.
\subsection{Query Evaluation}
\label{sec:backChain}

\newcommand{\solveByElim}{\code{solve\_by\_elim}}

As discussed in Sec.~\ref{sec:rules}, a query in the DSL is rewritten into a Lean theorem.
The type of the theorem is the Lean term corresponding to the query atom.
For the Lean type checker to accept the definition of this theorem, a value of that theorem type has to be assigned to the theorem object.
This value is effectively the proof of the theorem.

Backward-chaining (Sec.~\ref{sec:background}) is the algorithm typically used to construct this proof.
In Lean, when an algorithm is known to find proofs of a class of theorems, this algorithm is usually implemented as a tactic.
The tactic is expected to construct a proof term that is assigned to the theorem object.
The Lean type checker checks whether the constructed proof term is of the right type (the theorem expression) or not.
If it doesn't, Lean rejects the theorem definition.

The \solveByElim~tactic\footnote{\url{https://leanprover-community.github.io/mathlib4_docs/Lean/Meta/Tactic/SolveByElim.html}}, which is a part of the standard set of Lean 4 tactics, implements backward-chaining.
Maximum backtracking depth is one of the parameters of \solveByElim. 
If it is not specified by the user, a default value of 6 is used.
In addition, a set of theorems and/or axioms is another parameter.
Those are the theorems/axioms used by the tactic to try to prove the goal.
Instead of passing all those theorem/axiom names to the tactic, they can be annotated with a user-defined attribute, and that attribute is passed to \solveByElim. 

In Table~\ref{fig:rules}, the rewrite rule for queries assigns 
\begin{lstlisting}
by intros; solve_by_elim using datalog_rule    
\end{lstlisting}
to the theorem object generated for the query.
The Lean keyword \code{by} evaluates all subsequent code in tactic mode.
The \code{intros} tactic introduces any hypothesis or universally quantified variables into the proof context.
The semicolon after \code{intros} is a tactic combinator, evaluating its right-hand side only after the left-hand side succeeds.
In this case using this combinator is safe because the \code{intros} tactic always succeeds.
Now the \solveByElim~tactic is called, which takes an attribute name after the \code{using} tactic-specific keyword.
Since all the facts, rules, and used Lean theorems are already annotated with the \code{datalog\_rule} attribute, this is the one passed to the tactic.

\section{Examples}
\label{sec:examples}

\newcommand{\ra}{\code{r$_1$}}
\newcommand{\rb}{\code{r$_2$}}
\newcommand{\xa}{\code{.x$_1$}}
\newcommand{\ya}{\code{.y$_1$}}
\newcommand{\xb}{\code{.x$_2$}}
\newcommand{\yb}{\code{.y$_2$}}

This section illustrates using the Lean Datalog DSL on three examples: reachability, overlapping rectangles, and computing function derivatives.

\subsection{Reachability}

\begin{figure}[t]
\begin{lstlisting}[xleftmargin=10pt,firstline=3,lastline=17,numbers=left]
inductive edge: String → String → Prop
inductive path: String → String → Prop

datalog {

r1: path(x,y) :- edge(x,y).
r2: path(x,y) :- path(x,z), edge(z,y).
f1: edge("a", "b").
f2: edge("b", "c").
f3: edge("b", "d").

q0: path("a", "c")?
q1: path("b",m?)?
--q2: path("c", "d")? --failing
}
\end{lstlisting}
    \caption{Reachability example.}
    \label{fig:reachability}
\end{figure}

Node reachability in a graph is considered one of the most straightforward Datalog examples.
This is the same example we used in Sec.~\ref{sec:background} (Fig.~\ref{fig:datalogExample} and Fig.~\ref{fig:leanExample}), but now implemented in the Datalog DSL together with Lean.
The code in Fig.~\ref{fig:reachability} shows how Lean inductive predicates can be used together with Datalog rules, facts, and queries to prove the existence of a path between two nodes in a graph.
A graph node is identified by a \code{String} label.
Lines 1-2 are Lean inductive predicates defining edge and path relations. 
Each of the two predicates is empty (with no constructors).
Within the Datalog DSL section that comes next, those two relations are populated using Datalog facts and rules.

In lines 6-7, rules \code{r1} and \code{r2} recursively define a path within a graph. 
There is a path between \code{x} and \code{y} if there is an edge between \code{x} and \code{y} (rule \code{r1}), or if there is a path between \code{x} and \code{z}, and an edge between \code{z} and \code{y} (rule \code{r2}).
In lines 8-10, three facts (\code{f1}, \code{f2}, and \code{f3}) are defined, one fact for each edge in the graph. 

The first query in line 12 (\code{q0}) checks whether there is a path between \code{"a"} and \code{"c"}.
This query is proven using the Lean axioms generated by the facts and rules defined earlier.
The DSL interpreter formulates this query as a theorem with the same name \code{q0}, and tries to prove it using the rules and facts defined earlier.
The proof found using the backward chaining tactic in Lean is assigned to theorem \code{q0} as follows:
\begin{lstlisting}
    theorem q0 : path "a" "c" :=
       r2 (r1 f1) f2
\end{lstlisting}

The second query in line 13 (\code{q1}) includes a free variable \code{m?}, which is to be assigned a value by the backward chaining algorithm, together with Lean unification.
The implementation is expected to find an instantiation of variable \code{m?}, which is the endpoint of a path starting at node \code{"b"}.
In this example two such values exist: \code{"c"} and \code{"d"}.
Backward chaining is expected to arbitrarily find one satisfying assignment.
If the value found by the implementation is \code{"d"}, the following theorem together with its proof are assigned to \code{"q1"}.
\begin{lstlisting}
    theorem q1 : path "b" "d" :=
       r1 f3
\end{lstlisting}

The last query in line 14 (\code{q2}) is commented out, because it is not satisfiable given the rules and facts in this example.
A path from \code{"c"} to \code{"d"} is infeasible, so no proof can be found for the query theorem.
As a result, the Lean type system would reject this theorem.

\subsection{Overlapping Rectangles}
\label{sec:recOverlap}

\begin{figure}[t]
\begin{lstlisting}[xleftmargin=0pt,firstline=9,lastline=34,numbers=left]
structure Rect where
  x₁: ℤ
  y₁: ℤ
  x₂: ℤ
  y₂: ℤ

inductive intersect: Rect → Rect → Prop

def rect₁ := Rect.mk 50 50 400 100

def rect₂ := Rect.mk 75 25 125 300

def rect₃ := Rect.mk 150 80 425 200

datalog {

overlap: intersect(r₁, r₂) :- 
    (r₂.y₂ ≥ r₁.y₁), 
    (r₂.y₁ ≤ r₁.y₂), 
    (r₂.x₂ ≥ r₁.x₁), 
    (r₂.x₁ ≤ r₁.x₂).

q0: intersect(rect₁, rect₂)?
q1: intersect(rect₁, rect₃)?
-- failure q2: intersect(rect₂, rect₃)?
}
\end{lstlisting}
    \caption{Overlapping rectangles example.}
    \label{fig:overlap}
\end{figure}

\begin{figure}[t]
    \includegraphics[width=0.4\textwidth]{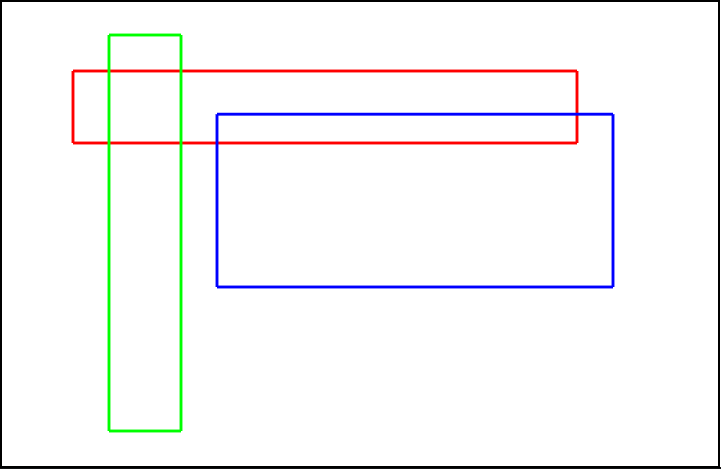}
    \caption{The three rectangles used in the overlapping rectangles example, with coordinates \textcolor{red}{\code{rect}$_1$: (50, 50, 400, 100)}, \textcolor{green}{\code{rect}$_2$: (75, 25, 125, 300)}, and \textcolor{blue}{\code{rect}$_3$: (150, 80, 425, 200)}.}
    \label{fig:rectangles}
\end{figure}

Given two rectangles in a 2-dimensional integer coordinate space (with the top-left corner at coordinate (0,0)), this example checks whether the two rectangles overlap (Fig.~\ref{fig:overlap}), and if they do, a proof is generated in the process.
In lines 1-5, a Lean structure \code{Rect} is defined, where a rectangle is defined in terms of its upper-left (\code{x$_1$,y$_1$}) and lower-right (\code{x$_2$,y$_2$}) coordinates.
In line 7, a binary inductive predicate \code{intersect} is then defined on two \code{Rect} values.
The three rectangles used in the example (lines 9-13) are plotted in Fig.~\ref{fig:rectangles}.
The rectangle objects are created using the \code{Rect.mk} constructor, which is automatically created by Lean for the \code{Rect} structure.
In this example, \code{rect$_1$} overlaps with both \code{rect$_2$} and \code{rect$_3$}, while \code{rect$_2$} and \code{rect$_3$} do not overlap with each other.

The most straightforward way to check if two rectangles overlap is to enumerate the cases where they \emph{do not} overlap, and negate the disjunction of those cases.
Given two rectangles \ra~and \rb, they do not overlap if 
(1) \rb~is completely above \ra~(\rb\yb~$<$~\ra\ya), 
(2) \rb~is completely below \ra~(\rb\ya~$>$~\ra\yb), 
(3) \rb~is completely to the left of \ra~(\rb\xb~$<$~\ra\xa), or 
(4) \rb~is completely to the right of \ra~(\rb\xa~$>$~\ra\xb).
The Datalog rule in lines 17-21 is the negation of the disjunction of those four cases.

Query \code{q0} (line 23) checks whether \code{rect$_1$} and \code{rect$_2$} overlap.
Similarly, query \code{q1} (line 24) checks whether \code{rect$_1$} and \code{rect$_3$} overlap.
The proof of \code{q0} involves several definitions from the Lean library of integers that are not relevant to this discussion.
The proof states that rule \code{overlap} was used, with four sub-proofs of the four premises of that rule. 
\begin{lstlisting}[xleftmargin=10pt,firstline=38,lastline=42]
theorem q0 : intersect rect₁ rect₂ :=
   overlap (Int.NonNeg.mk (300 - Nat.succ 49)) 
           (Int.NonNeg.mk (100 - Nat.succ 24)) 
           (Int.NonNeg.mk (125 - Nat.succ 49))
           (Int.NonNeg.mk (400 - Nat.succ 74))
\end{lstlisting}

On the other hand, query \code{q2} in line 25 tries to certify that rectangles \code{rect$_2$} and \code{rect$_3$} overlap. Since those two rectangles are non-overlapping, this query is not satisfiable, so it is commented out.

\subsection{Computing Function Derivatives}
\label{sec:derivatives}

The Mathlib~\cite{Mathlib:2020} Lean library of mathematical definitions and theorems includes a formalization of derivatives of continuous mathematical functions as a part of the Calculus package\footnote{Mathlib.Analysis.Calculus.Deriv}.
Because of the compositional nature of function derivatives, the theorems in this formalization can be collectively used to prove the correctness of arbitrary function derivates.
This example leverages both Mathlib derivative theorems and Lean unification to compute the derivates of functions defined on real numbers.

Mathlib defines a ternary predicate \code{HasDerivAt} of type 
\begin{lstlisting}
    (f: F → G) → (f' : G) → (x : F) → Prop
\end{lstlisting}
where F and G are arbitrary types with some typeclass constraints (more details in Mathlib documentation 
\footnote{\url{https://leanprover-community.github.io/mathlib4_docs/Mathlib/Analysis/Calculus/Deriv/Basic.html\#HasDerivAt}}).
Given a function \code{f} with domain \code{F} and range \code{G}, \code{HasDerivAt} asserts that given \code{x} $\in$ \code{F}, the derivative of \code{f} at \code{x} is \code{f'}.
For example, the Mathlib theorem \code{hasDerivAt\_sin} is of type
\begin{lstlisting}
    hasDerivAt_sin (x : ℝ) : HasDerivAt sin (cos x) x
\end{lstlisting}
stating that for all real values \code{x}, the derivative of \code{sin x} is \code{cos x}.

\begin{figure}[t]
\begin{lstlisting}[xleftmargin=0pt,firstline=35,lastline=44,numbers=left]
def drv: (ℝ → ℝ) → (ℝ → ℝ) → Prop := 
   fun f f' ↦ ∀x:ℝ, HasDerivAt f (f' x) x

datalog {

use hasDerivAt_sin, hasDerivAt_cos, HasDerivAt.add.

q0: drv(Real.sin, h?)?
q1: drv(fun x ↦ Real.cos x + Real.sin x, g?)?
}
\end{lstlisting}
    \caption{Function differentiation example.}
    \label{fig:diff}
\end{figure}

Figure~\ref{fig:diff} starts with a Lean binary predicate \code{drv} (lines 1-2) that wraps \code{HasDerivAt}, abstracting the value \code{x} into the domain type of the function \code{f} and its derivative \code{f'}.
The Datalog section starts with a \code{use} clause (line 6), adding existing Lean theorems as Datalog Rules.
The three Mathlib theorems used here formalize the derivatives of \code{Real.sin}, \code{Real.cos}, and the addition operator.

Query \code{q0} at line 8 aims at finding the derivative of the \code{Real.sin} function. The query definition refers to the derivative with the placeholder \code{h?}. The backward chaining algorithm tries to use the three Mathlib theorems listed previously in the \code{use} statement, unifying \code{h?} with \code{Real.cos} as indicated in the type of \code{hasDerivAt\_sin}.
Evaluating the query results in this proof:
\begin{lstlisting}
    theorem q0 : drv sin cos :=
        hasDerivAt_sin
\end{lstlisting}

Query \code{q1} at line 9 tries to find the derivative of the function \code{$\lambda$ x $\mapsto$ Real.cos x + Real.sin x}. 
This function composes real number addition, \code{Real.sin}, and \code{Real.cos}.
Backward chaining composes the use of different theorems, computing the derivative function \code{$\lambda$ x $\mapsto$ -sin x + cos x}, and generating this proof:
\begin{lstlisting}
    theorem q1 : 
        drv (fun x => cos x + sin x) 
            fun x => -sin x + cos x :=
        fun x => HasDerivAt.add 
                    (hasDerivAt_cos x) 
                    (hasDerivAt_sin x)
\end{lstlisting}
\section{Related Work}
\label{sec:related}

Both declarative logic programming languages (e.g., Datalog) and Interactive Theorem Provers (ITPs) have been ubiquitously used in various application domains.
This section focuses on related work integrating the use of both Datalog and ITPs in different ways, in addition to use cases of both technologies in reasoning about software systems.

Lean has been used to certify different aspects of software systems, and in various software analysis and verification projects recently.
Examples include formalizing authorization policies and access permissions in Cloud-based systems~\cite{Cutler:2024}, the certified analysis of assurance cases of automotive software~\cite{Shahin:2021}, and the certified construction of safety cases~\cite{Viger:2021}.

Datalog has also been extensively used in declarative reasoning about software systems. 
Examples include security analysis of web applications~\cite{Lam:2005}, pointer analysis~\cite{Bravenboer:2009} and taint analysis~\cite{Grech:2017} of heap manipulating programs, analysis of interrupt-driven software~\cite{Sung:2017}, flow-sensitive analysis of concurrent programs~\cite{Kusano:2016,Kusano:2017}, analysis of variability-aware programs~\cite{Shahin:2022}, and analysis of automotive software product lines~\cite{Shahin:2023},

ITPs have been used to reason about the semantics of Datalog and its inference algorithms.
Datalog with stratified negation was formalized in Coq and SSReflect~\cite{Benzaken:2017}.
As a part of this project, Coq was used to prove the correctness and termination of fixed point Datalog semantics.
A couple of Datalog evaluation optimizations were also certified in Coq, using a formalized trace semantics of Datalog~\cite{Begay:2021}.
Lean was also used to verify Datalog evaluation algorithms~\cite{Tantow:2025}.
These are all examples of using an ITP to prove some properties of Datalog semantics and inference algorithms.
This paper on the other hand does not model or reason about the semantics of Datalog syntactic structures, but instead, Datalog constructs are translated directly into Lean, and used as a proof search technique.

A different research track focuses on integrating Datalog into other programming languages and formalisms, or adding programming constructs from other paradigms to Datalog.
For example, Datafun~\cite{Arntzenius:2016,Arntzenius:2019} adds higher-order functional programming to Datalog while maintaining monotonicity of inference.
Formulog~\cite{Bembenek:2020} extends Datalog with first-order functional programming constructs, facilitating the smooth interoperability between the Datalog engine and SMT solvers.
Flix~\cite{Madsen:2016} is a language integrating Datalog constructs into an imperative language, particularly to allow declarative reasoning on lattice structures.

This paper is similar to those projects in the sense that it is an integration of Datalog syntax into another language.
One major difference though is that our integration is \term{light-weight}, and implemented as a layer of syntactic sugar on top of standard Lean syntax and semantics.
This leverages the logical aspects of Lean, particularly inductive predicates, theorems as types, proof tactics, and dependent types.
Another implementation difference is how our Datalog integration utilizes Lean metaprogramming to add Datalog syntax as a non-intrusive DSL.

This project also slightly extends Datalog in two ways: allowing arbitrary Lean terms in field values, and using Lean theorem definitions as Datalog rules.
Many other projects extend Datalog in various ways, including probabilistic Datalog~\cite{Fuhr:1995}, variability-aware Datalog~\cite{Shahin:2020}, and differential Datalog for incremental computation~\cite{Ryzhyk:2019}.
Modern Datalog engines such as \souffle~\cite{Jordan:2016} also extend Datalog with a richer type system, including records, Algebraic Data Types (ADTs), and subtypes.
\section{Conclusion and Future Work}
\label{sec:conclusion}

This paper presented the design and implementation of a Datalog Domain Specific Language (DSL) embedded in the Lean Interactive Theorem Prover (ITP).
The implementation of this DSL leverages different Lean metaprogramming facilities such as syntax extensions, user-defined elaborators, and tactic-based proofs.
A primary goal of the DSL design is bidirectional interoperability between the Datalog DSL and Lean. 
In addition to rules and facts, Datalog queries are automatically translated into Lean theorems, and tactic-based backward-chaining is used to compute their proofs. 

The paper also presents three examples illustrating the use of the DSL.
The examples cover query evaluation using backward-chaining, using existing Lean definitions and theorems from within Datalog constructs, the automatic generation of proof trees for goals in the form of Lean proof terms, and filling-in some of the missing values in goals using leveraging the Lean type-based unification.

This DSL is only a first step towards adding declarative proof search facilities to Lean.
Several query evaluation optimizations already used in Datalog engines can be added to this DSL implementation.
Another potential future direction is adding forward-chaining, but only allowing its use in programs where all Datalog relations are defined over finite types.
Without this restriction, the termination of forward-chaining is not guaranteed.

\begin{acks}
Thanks to the anonymous reviewers of this paper for their thorough feedback and insightful comments and suggestions.  
\end{acks}

\bibliographystyle{ACM-Reference-Format}
\bibliography{datalog,autodiff,pl,logic}
\end{document}